\documentclass[twocolumn,aps,floats,superscriptaddress,prl]{revtex4} 

\usepackage[english]{babel}
\usepackage{amsmath}
\usepackage{amssymb}
\usepackage{graphicx}

\newcommand{\pnu}[1] {\overset{\smash{\scriptscriptstyle (-)}}{\nu}_{\hskip-3pt #1}}
\newcommand{\eps}{\epsilon_{e\tau}}
\newcommand{\si}{S_{13}}

\begin{document}
\preprint{IFIC 01/62}
\preprint{TUM-HEP 444/01}
\preprint{MPI-Pht/2001-48}

\title{How sensitive is a neutrino factory to the angle $\theta_{13}$?}

\author{P. Huber}
\email{phuber@ph.tum.de}
\affiliation{Instituto de F\'{\i}sica Corpuscular -- C.S.I.C., 
Universitat de Val{\`e}ncia 
Edificio Institutos, Aptdo.\ 22085, E--46071 Val{\`e}ncia, Spain\\}
\affiliation{Theoretische Physik, Physik Department, 
Technische Universit{\"a}t M{\"u}nchen,
James--Franck--Strasse, D--85748 Garching, Germany}
\affiliation{Max-Planck-Institut f{\"u}r Physik, Postfach 401212, 
D--80805 M{\"u}nchen, Germany}
\author{T.~Schwetz}
\email{schwetz@ific.uv.es}
\affiliation{Instituto de F\'{\i}sica Corpuscular -- C.S.I.C., 
  Universitat de Val{\`e}ncia 
Edificio Institutos, Aptdo.\ 22085, E--46071 Val{\`e}ncia, Spain\\}
\author{ J.~W.~F. Valle}
\email{valle@ific.uv.es}
\affiliation{Instituto de F\'{\i}sica Corpuscular -- C.S.I.C., 
  Universitat de Val{\`e}ncia 
Edificio Institutos, Aptdo.\ 22085, E--46071 Val{\`e}ncia, Spain\\}

\date{19.11.2001}

\begin{abstract}
  
  We consider the impact of non-standard interactions of neutrinos
  (NSI) on the determination of neutrino mixing parameters at a
  neutrino factory using $\pnu{e}\to\pnu{\mu}$ ``golden channels'' for
  the measurement of $\theta_{13}$. We show how a small residual NSI
  leads to a drastic loss in sensitivity in $\theta_{13}$, of up to
  two orders of magnitude. This can be somewhat overcome if two
  baselines are combined.

\end{abstract}
\maketitle

The long-standing solar and atmospheric neutrino
anomalies~\cite{solatmos} now give compelling evidence for new physics
in the lepton sector.  The data are well described in the simplest
3-neutrino oscillation framework~\cite{Gonzalez-Garcia:2001sq} defined
by two mass-squared differences $\Delta m^2_\mathrm{sol}$ and $\Delta
m^2_\mathrm{atm}$, three angles $\theta_{12}$, $\theta_{23}$,
$\theta_{13}$ and one relevant CP-violating phase $\delta$. The
parameters $\Delta m^2_\mathrm{atm}$ and $\theta_{23}$ are determined
by atmospheric experiments, while $\Delta m^2_\mathrm{sol}$ and
$\theta_{12}$ are determined by solar experiments.
For $\theta_{13}$ there is only an upper bound $\sin^2 2\theta_{13}
\lesssim 0.1$ at 90\%~CL derived from reactor
experiments~\cite{chooz}.  The phase $\delta$ is completely unknown.

A new generation of long-baseline neutrino oscillation experiments
using a neutrino beam from the decay of muons in a storage ring is
discussed~\cite{nuFacPapers}.  Such so-called \texttt{neutrino
  factories} should give a precise determination of $\theta_{13}$,
possibly down to few$\times 10^{-4}$, exploring also the
important issue of leptonic CP violation~\cite{Dick}.

In a large class of models neutrino masses are accompanied by
non-standard interactions of neutrinos (NSI). These can be
non-universal (NU) or flavour-changing (FCI), and may arise also in
the absence of neutrino mass~\cite{NSImodels}. Such non-standard
interactions of neutrinos affect their propagation in matter, with
implications for solar
\cite{Valle:1987gv,MSW,NSIrecent,Bergmann:2000gp},
atmospheric~\cite{Gonzalez-Garcia:1999hj,Val}, and astrophysical
neutrino sources~\cite{NSIastro}.
The possibility to use a neutrino factory for probing NSI with the
earth matter, which neutrinos have to cross necessarily in a
long-baseline experiment, has been previously examined in the
appearance channels $\pnu{e}\to\pnu{\tau}$ \cite{Gago:2001xg} and
$\pnu{\mu}\to\pnu{\tau}$ \cite{Gago:2001xg,Huber:2001zw}.
 
In this letter we will consider the impact of NSI on the determination
of neutrino mixing parameters assuming the standard setup of a
neutrino factory.  Therefore, we will focus on the
$\pnu{e}\to\pnu{\mu}$ channels, i.~e.  the ``golden channels'' for the
measurement of $\theta_{13}$.  In contrast to
refs.~\cite{Gago:2001xg,Huber:2001zw} we need only to assume a muon
detector, which seems more realistic than a tau detector.  We treat
NSI and neutrino oscillations in a common framework to investigate
correlations between the corresponding parameters. We show that the
presence of a small FCI in the $e-\tau$ channel, safely within current
limits, leads to a drastic loss in sensitivity in $\theta_{13}$, up to
two orders of magnitude. This follows from a strong correlation
between these parameters.  We show how this effect can be diminished
to some extent if two baselines are combined.

In the presence of NSI the evolution of neutrinos passing through
matter is modified by a non-standard potential due to the coherent
forward scattering amplitude of NSI-processes $\nu_\alpha + f \to
\nu_\beta + f$, where $f$ is a fermion in the medium. For simplicity
we will consider only NSI with down-quarks: $f=d$. Then the potential
induced by NSI can be written as $V_\mathrm{NSI} = \sqrt{2} G_F
N_d\,\epsilon_{\alpha\beta}$, where $N_d$ is the number density of
down-quarks along the neutrino path.  The small numbers
$\epsilon_{\alpha\beta}$ describe the NSI in units of the
Fermi-constant $G_F$. The off-diagonal elements
$\epsilon_{\alpha\beta}$ with $\alpha\neq\beta$ correspond to FCI,
whereas the differences in the diagonal elements
$\epsilon_{\alpha\alpha}$ lead to NU interactions.  For simplicity we
take all $\epsilon_{\alpha\beta}$ real. Present data on various lepton
flavour violating processes can be used to set upper bounds on
FCI~\cite{Bergmann:2000gp,BergmannLSND,BergmannBounds} leading to
$|\epsilon_{e\mu}| \lesssim 7\times 10^{-5},\: |\epsilon_{\mu\tau}|
\lesssim 5\times 10^{-2},\: |\epsilon_{e\tau}| \lesssim 7\times
10^{-2}$.  Much weaker bounds of order 0.1 apply for the NU
coefficients.

In this letter we show how claimed sensitivities on $\theta_{13}$ get
destroyed if a small NSI component exists. For this it suffices to
consider a simplified scenario in which we neglect the solar
mass-squared difference. If we parameterize the lepton mixing matrix
$U_{23}U_{13}U_{12}$, this implies also that the angle $\theta_{12}$
and the phase $\delta$ disappear. Further, we restrict ourselves to
the elements $\epsilon_{e\tau}$ and $\epsilon_{\tau\tau}$ (the case
$\epsilon_{\mu\tau}$ and $\epsilon_{\tau\tau}$ is considered in
Ref.~\cite{Huber:2001zw}), where present constraints are rather weak.
Then the evolution of flavour neutrinos is governed by the Hamiltonian
\begin{eqnarray}\label{ham2}
H_\nu &=& 
U_{23} U_{13}\,\mbox{diag}\,(0,0,\Delta)\,U^\dagger_{13}U^\dagger_{23}\nonumber\\
&&+
V\,\left(\begin{array}{ccc} 
1 & 0 & r\eps \\ 0&0&0 \\ r\eps & 0 & r\epsilon_{\tau\tau}
\end{array}\right)\,. 
\end{eqnarray}
Here we have defined $\Delta:=\Delta m^2_\mathrm{atm}/2E_\nu$ and
$V=\sqrt{2}G_F N_e$ is the matter potential due to the Standard Model
charged current interaction~\cite{MSW}, where $N_e$ is the electron
number density and $r:= V_\mathrm{NSI}/V=N_d/N_e$ with $r\approx
3$ in earth matter. In the Hamiltonian (\ref{ham2}) a sign change of
$\Delta$ is equivalent to a sign change of $V$, which interchanges the
evolution of neutrinos and anti-neutrinos.  In the following we show
that even in this simple scenario the strong correlation between
$s_{13}$ and $\eps$ will lead to serious complications for the
measurement of $\theta_{13}$ at a neutrino factory.

To get an intuitive understanding of the numerical results it is
useful to derive first an analytic expression for the appearance
probability.  To this aim we assume a constant matter potential $V$
and, since $s_{13}$ must be small, consider all terms containing
$s_{13},\,\eps$ and $\epsilon_{\tau\tau}$ as a small perturbation of
the Hamiltonian and calculate eigenvalues and eigenvectors of
eq.~\ref{ham2} up to first order in these small quantities. The
resulting appearance probability is second order and has the form
\begin{equation}\label{prob}
P_{\nu_e\to\nu_\mu}\approx A\,s_{13}^2 + B\,s_{13}\eps + C\,\eps^2 \,.
\end{equation}
where the coefficients $A,\,B$ and $C$ are of the same order and
depend on the neutrino energy, on the baseline $L$ and on the sign of
$V$ (neutrinos or anti-neutrinos) in a nontrivial way. For example,
\begin{eqnarray}
\label{coeffs}
A &=& 4\, s_{23}^2 \left(\frac{\Delta}{\Delta-V}\right)^2
      \sin^2\frac{(\Delta-V)L}{2}
\end{eqnarray}
with similar expressions for $B$ and $C$.  These analytic expressions
for the appearance probabilities are in agreement with numerical
calculations within a few \% in the relevant parameter range.  We note
that the NU coefficient $\epsilon_{\tau\tau}$ does not show up in
leading order in the appearance probability \cite{Gago:2001xg}.
Therefore, we will neglect it in our numerical studies and concentrate
only on FCI.
  We consider both signs for $\eps$, this sign however becomes irrelevant as $s_{13}$ approaches
zero.
 \begin{figure}[htb!]
  \begin{center}
    \includegraphics[width=0.35\textwidth]{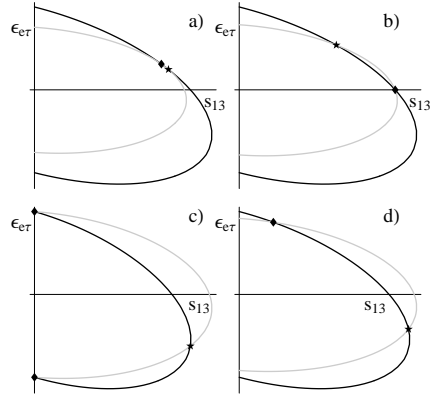}
    \caption{Lines of constant event rates in 
      $s_{13}$--$\epsilon_{e\tau}$ (black for neutrinos and grey for
      anti-neutrinos) at $L=3\,000\,\mathrm{km}$.}
    \label{fig:ellipses}
  \end{center}
\end{figure}

Eq.~\ref{prob} implies that the lines of constant event rates in the
$s_{13}$--$\epsilon_{e\tau}$ plane are ellipses, whose coefficients
are obtained by fixing a baseline and folding $A,B$ and $C$ with 
flux and detection cross section.  In fig.~\ref{fig:ellipses} we
draw these ellipses for neutrinos and anti-neutrinos through different
start points marked with a star. In general the two ellipses will
cross not only at the start point. All points where the ellipses cross
will lead to identical total event rates and are therefore a source of
possible confusion. Depending on the choice of initial values four
typical cases can be distinguished.  Case (a) has two solutions witch
are very close and away from both axes.  In this case both NSI and
oscillations can be established.  Case (b) has a second solution on
the $\eps=0$ axes. Therefore one can only establish the existence
of oscillations.  Case (c) has additional solutions on the $s_{13}=0$
axes, which means that only NSI can be established but not
oscillation. Case (d) has solutions which are close to both axes.  If
the count rate is not high enough the allowed region will touch both
axes, leading to complete confusion between NSI and oscillation.  In
the following we explore this through a detailed numerical example and
show that all these four cases can actually arise for realistic
oscillation and NSI parameter choices.

In our numerical calculations we assume a neutrino factory with an
energy of 50 GeV for the stored muons and $2\times 10^{20}$ useful
muon decays of each polarity.  We consider a magnetized iron detector
with perfect charge identification of the muons and a mass of
$10\,\mathrm{kt}$.  The muon detection threshold is set to 4 GeV, the
energy resolution of the detector is approximated by a Gaussian
resolution function with $\Delta E_\nu/E_\nu=10\%$ and we use 20 bins
in muon energy.  The appearance probability is obtained by numerically
solving the neutrino evolution equation with the Hamiltonian
(\ref{ham2}), using a realistic earth matter density profile.  Then
this probability is folded with neutrino flux, cross section and
energy resolution function to obtain the event rates in the detector,
following~\cite{Freund:2001ui}.

We consider only the $\nu_e\to\nu_\mu$ appearance channel since it
contains most of the information.  Our ``observables'' are the event
rates of this channel $n^i_\nu \:(n^i_{\bar\nu})$ for (anti-)neutrinos
in each energy bin $i$. We fix the atmospheric oscillation parameters
at $\Delta m^2_\mathrm{atm}= 3\times 10^{-3}\,\mathrm{eV}^2$ and
$\sin^22\theta_{23}=1$. Thus, at a given baseline the event rates
depend on $\si := \sin^22\theta_{13}$ and $\eps$.
In order to evaluate the potential of a neutrino factory to probe
these parameters we proceed as follows. For any given
$(\si^0,\,\eps^0)$ choice we calculate the event rates
$n_x^i(\si^0,\,\eps^0)$ with $x=\nu,\,\bar\nu$ and
construct a $\chi^2(\si,\eps;\, \si^0,\eps^0)$ appropriate for a
Poisson distribution.
Using this we derive allowed regions in the $\si - \eps$ plane
as usual. For
all $\si$ and $\eps$ values inside these confidence regions the data
can be fitted at the given CL (in our case 99\%) if the true values 
are $\si^0$ and $\eps^0$. Thus, these regions indicate the capability 
of a neutrino factory to probe $\si$ and $\eps$.

\begin{figure}[htb!]
  \begin{center}
    \includegraphics[width=0.35\textwidth]{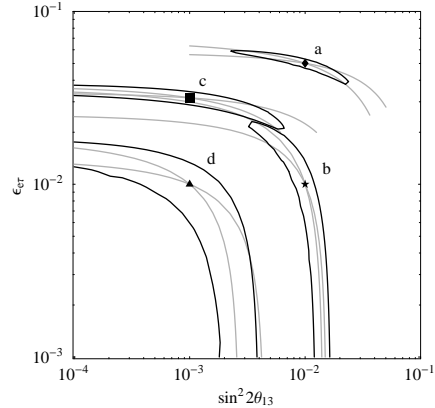}
    \caption{99\% CL allowed regions (black lines) in the $\sin^2 2\theta_{13}$--$\epsilon_{e\tau}$ 
      plane for different input values, as indicated by the points, at
      a baseline of $3\,000\,\mathrm{km}$. Lines of constant event
      rates are displayed in grey.}
    \label{fig:fits}
  \end{center}
\end{figure}

In fig.~\ref{fig:fits} we show such fits for four characteristic test
points $(\si^0,\eps^0)$ at a baseline of $3\,000\,\mathrm{km}$ in
correspondence with the cases considered in fig.~\ref{fig:ellipses}.
The two grey lines passing through each of these test points are the
lines of constant total rates for neutrinos and anti-neutrinos.  From
the fact that the confidence regions follow closely the lines of
constant rates we learn that most information is contained in the
total rates; the spectral information is not very important. We have
verified that our results are, indeed, rather insensitive to
variations of the number of energy bins and of the energy resolution.
Moreover we find that one must take into account simultaneously both
neutrino and anti-neutrino rates (this requires to run the neutrino
factory in both polarities).  This is easy to understand since the
allowed regions extend as long as the lines of constant neutrino and
anti-neutrino rates remain close to each other, i.~e.~\texttt{both}
rates are similar to the ones at the test point.  In regions where the
lines are far from each other the neutrino or the anti-neutrino rates
are very different from the one in the test point, and hence it is not
possible to fit the data generated by the parameters $(\si^0,\eps^0)$
in those regions.

Let us consider in more detail the fits for the four different test
points in fig.~\ref{fig:fits}.
For case (a) with large $\si^0$ and $\eps^0$ values one would need
both oscillations and FCI.
In case (b) ($\si^0 = 10^{-2},\, \eps^0=
10^{-2}$) the allowed region extends to the $\eps=0$ axis.  This
means that the data in a neutrino factory generated by these
parameters require oscillations.
In contrast, for case (c) ($\si^0 =
10^{-3},\, \eps^0= 3.2 \times 10^{-2}$) the allowed region extends only
to the $\si=0$ axis, indicating that the data in a neutrino factory
generated by these parameters require the presence of FCI.
Finally, in case (d) ($\si^0 = 10^{-3},\, \eps^0=
10^{-2}$) there is complete confusion between oscillations and FCI,
because the allowed region extends to both axes $\eps=0$ and $\si=0$.
This means that the data in a neutrino factory generated by these
parameters can be fitted either with pure oscillations ($\eps=0$) or
pure FCI ($\si=0$) with both signs of $\eps$.
We can define a sensitivity to $\si$ by requiring that the allowed
region for a given point in the $\si-\eps$ plane does not touch the
$\si=0$ axis. By scanning the $\si-\eps$ plane we obtain the following
sensitivity limits for three baselines: 
\begin{center}
\begin{tabular}{c|ccc}
baseline&$700$&$3000$&$7000$\\
\hline
w/o NSI&$2 \times 10^{-4}$&$3 \times 10^{-4}$&$5 \times 10^{-4}$\\
w NSI&$1 \times 10^{-1}$&$7 \times 10^{-2}$&$5 \times 10^{-3}$\\
\end{tabular}
\end{center}
Comparing with the sensitivity limits without NSI~\cite{nuFacPapers}
we can see that the sensitivity loss is dramatic, especially for
shorter baselines.
 \begin{figure}[htb!]
  \begin{center}
    \includegraphics[width=0.35\textwidth]{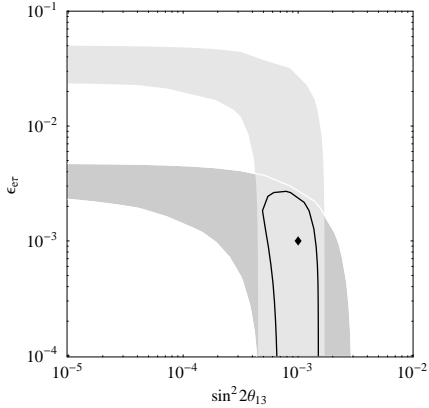}
    \caption{$\sin^2 2\theta_{13}$--$\epsilon_{e\tau}$ sensitivities 
when combining two baselines at 700 km, at 7000 km, see text.}
    \label{fig:sens}
  \end{center}
\end{figure}

The geometry of the muon storage rings currently discussed offers the
striking possibility to illuminate two detectors at different
baselines with neutrinos from one neutrino factory.  In fig.~3 we show
the 99\% CL region from a fit to data generated from the test point
$\si^0=10^{-3}$ and $\eps^0=10^{-3}$ at 700 km, at 7000 km and from a
combination of both. In this particular case, for each of the two
baselines taken alone there is complete confusion between $\si$ and
$\eps$. However, combining the two baselines we obtain a much smaller
allowed region indicated by the solid line, which is roughly given by
the intersection of the individual regions.  Here the data can be
fitted with $\eps=0$, with improved sensitivity for $\si$.
For the combination of two baselines the  analysis is
performed as before, but using the sum of the $\chi^2$-functions for
the individual baselines. 
We have evaluated neutrino factory sensitivities (defined as above)
for all possible combinations of the three baselines 700 km, 3000 km
and 7000 km.
\begin{center}
\begin{tabular}{c|ccc}
baseline&$700\& 3000$&$700\& 7000$&$3000\& 7000$\\
\hline
w/o NSI&$2 \times 10^{-4}$&$3 \times 10^{-4}$&$3 \times 10^{-4}$\\
w NSI&$2 \times 10^{-3}$&$2 \times 10^{-3}$&$2 \times 10^{-3}$\\
\end{tabular}
\end{center}
The sensitivities for $\si$ are still considerably worse (up to one
order of magnitude) than those for the pure oscillation case (no NSI).
Notice that no substantial improvement is expected by changing the
energy of the stored muons since neutrinos carry a relatively wide
spectrum already for our nominal 50 GeV choice.

We have shown within a simplified scenario that the presence of even a
small NSI component, predicted in most models of neutrino mass, can
deteriorate the sensitivity of a neutrino factory for $\theta_{13}$ by
orders of magnitude. This problem can partly be resolved by combining
two baselines. Additional effects of NSI in neutrino source and
detection~\cite{source}, 
and CP violating phases in NSI coefficients, as well as in the mixing
matrix will lead to even more serious difficulties~\cite{huber}.

This work was supported by the European Commission grants
HPRN-CT-2000-00148 and HPMT-2000-00124, by the ESF \emph{Neutrino
  Astrophysics Network} and by Spanish MCyT grant PB98-0693.  We thank
discussions with H. Nunokawa whose work~\cite{Gago:2001xg} stimulated
our interest.

\end{document}